\documentclass[twocolumn]{aastex61}
\usepackage{amsmath,amstext}
\usepackage[T1]{fontenc}
\usepackage{apjfonts} 
\usepackage[figure,figure*]{hypcap}
\usepackage{booktabs}

\shorttitle{Multiple Spectral Components in the M84 jet}
\shortauthors{Meyer et al.}

\begin{document}

\title{Detection of an optical/UV jet/counterjet and Multiple Spectral Components in M84}

\author{Eileen T. Meyer}
\affiliation{Department of Physics, University of Maryland, Baltimore County, 1000 Hilltop Circle, Baltimore, MD 21250, USA}

\author{Maria Petropoulou}
\affiliation{Department of Astrophysical Sciences, Princeton University, 4 Ivy Lane, Princeton, NJ, 08544, USA}

\author{Markos Georganopoulos}
\affiliation{Department of Physics, University of Maryland, Baltimore County, 1000 Hilltop Circle, Baltimore, MD 21250, USA}
\affil{NASA Goddard Space Flight Center, Code 663, Greenbelt, MD 20771, USA}

\author{Marco Chiaberge}
\affiliation{Space Telescope Science Institute, 3700 San Martin Drive, Baltimore, MD 21218, USA}

\author{Peter Breiding}
\affiliation{Department of Physics, University of Maryland, Baltimore County, 1000 Hilltop Circle, Baltimore, MD 21250, USA}


\author{William B. Sparks}
\affiliation{Space Telescope Science Institute, 3700 San Martin Drive, Baltimore, MD 21218, USA}

\begin{abstract}
We report an optical/UV jet and counterjet in M84, previously unreported in archival HST imaging. With archival VLA, ALMA, and \emph{Chandra} imaging, we examine the first well-sampled spectral energy distribution of the inner jet of M84, where we find that multiple co-spatial spectral components are required. In particular, the ALMA data reveal that the radio spectrum of all four knots in the jet turns over at approximately 100 GHz, which requires a second component for the bright optical/UV emission. Further, the optical/UV has a soft spectrum and is inconsistent with the relatively flat X-ray spectrum, which indicates a third component at higher energies. Using archival VLA imaging, we have measured the proper motion of the innermost knots at 0.9$\pm$0.6 and 1.1$\pm$0.4$c$, which when combined with the low jet-to-counterjet flux ratio yields an orientation angle for the system of 74$^{+9}_{-18}$$^\circ$. In the radio, we find high fractional polarization of the inner jet of up to 30\% while in the optical no polarization is detected (< 8\%).  We investigate different scenarios for explaining the particular multi-component SED of the knots. Inverse Compton models are ruled out due to the extreme departure from equipartition and the unrealistically high total jet power required. The multi-component SED can be naturally explained within a leptohadronic scenario, but  at the cost of very high power in relativistic protons. A two-component synchrotron model remains a viable explanation, but more theoretical work is needed to explain the origin and properties of the electron populations. 
\end{abstract}

\section{Introduction}

Optical observations of resoloved extragalactic jets date back nearly one hundred
years, to the observation of the 20$''$-long jet in M87 made by by
\cite{curtis1918}. Several ground-based observations of jet knots
and hotspots (usually far from the host galaxy) have occurred in the
intervening years, though they remain relatively rare \citep[e.g.,][]{cayatte1987,fraix1988,falomo2009}. The
launch of the Hubble Space Telescope (HST), which provides observations with sub-arcsecond resolution and lower backgrounds, added significantly to the number of optically detected
jets, typically lower-power FR~Is\footnote{\cite{fan74} type I sources are typically low in radio power and have edge-darkened 'plumey' jet morphologies.  FR IIs correspond to powerful radio sources with highly collimated jets terminating in bright hot-spots, and are associated with higher Lorentz factors at the jet base \citep[e.g.,][]{kharb2010}.} \citep[e.g.,][]{sparks95_3c78,sparks1996_m87,martel1998_3c15, martel1999_jets,scarpa1999,floyd2006_3c293} where the synchrotron emission typically peaks at or just beyond the optical band. However, detections of resolved jets in the optical remain rare, with perhaps a few dozen detections in total.  This is likely due to a combination of lack of observations (the vast majority of thousands of known radio jets have never been observed with HST, or lack deep observations) and the particularities of the broad-band spectral energy distribution (SED) in some jets. In more powerful FR~IIs, for example, it seems that the synchrotron peak is well below the optical, based on deep upper limits \citep[e.g.,][]{tavecchio2007} as well as very soft spectra in deep observations \citep{sambruna2006,meyer2015,breiding2017}.  This, in combination with the competition from host galaxy light in many cases, leads to very low detection rates.  In consequence, the SEDs of most resolved jets are not measured at all beyond the radio, which means that it is difficult to assess theoretical models of jets on these scales, or to quantify e.g., the total energy they carry (with implications for jet feedback models).

With the launch of the \emph{Chandra} X-ray observatory in 1999 came the realization that many resolved radio jets produce significant X-ray emission, at kpc distances from the bright `core' at the base of the jet \citep[][and many more]{chartas2000,sambruna2002,sambruna2004,sambruna2006,harris2006,marshall2011,kharb2012}. Well over 150 sources have been detected, including both jets and terminal hotspots\footnote{The XJet Database at https://hea-www.harvard.edu/XJET/ contains most, but not all of these observations.}. In many cases, the X-ray emission from the kpc-scale jet is both hard and far above what is expected from the radio-optical synchrotron spectrum, indicating a second emission component, leading to their sometime designation as `anomalous' X-ray jets. For many years the preferred explanation for the anomalously high X-ray flux (found in mostly powerful FR~II sources) has been inverse Compton upscattering of the Cosmic Microwave Background (IC/CMB) by a jet which remains highly relativistic on kpc scales \citep{tavecchio2000,celotti2001}. However, this explanation predicts a well-defined, very high level of gamma-ray emission \citep{geo06} which has subsequently not been seen in several cases \citep{meyer2014,meyer2015,meyer2017,breiding2017}.  The model, which requires that the knots of the jets are moving `packets' of plasma, has also been ruled out in 3C~273 on the basis of stringent upper limits on the optical proper motions \citep{meyer2016}, and in PKS~1136-135 based on the high degree of polarization in the UV portion of the second component \citep{cara2013}.  The IC/CMB model is also at odds with a number of observed properties of \emph{Chandra}-detected jets, for example the observed variability \citep{marshall2010} and jet-to-counterjet flux ratio in Pictor~A \citep{hardcastle2016}, offsets between radio and X-ray emission and jet-to-counterjet flux ratio in 3C 353 \citep{kataoka2008}, as well as the offsets between radio and X-ray knots in 3C 111 \citep[][]{clautice2016}, among others.

With the IC/CMB model now in serious doubt, remaining alternatives to explain the X-rays are few.  In one of the papers announcing the discovery of the very first anomalous X-ray jet, PKS~0637$-$752, \cite{schwartz2000} ruled out synchrotron self-Compton and thermal bremsstrahlung as possible mechanisms to produce the X-rays, with the former requiring gross departures from equipartition and unrealistic beaming parameters to match the X-ray flux, and the latter requiring unreasonably high electron densities which exceed limits implied by rotation measure studies (a similar result is found in this paper and in \citealt{harris2002} for M84). However, a second synchrotron spectrum has long been considered a possible explanation for the X-rays \citep{atoyan2004,harris2004,kataoka2005,hardcastle2006,jester2006,uchiyama2006}, with the main detraction being the unexpected and unexplained additional electron energy distribution required. Hadronic emission models are also a possibility \citep{aharonian2002,kusunose2017,petropoulou2017}, though they have not been extensively explored in this context.

While the debate over the X-ray emission has unfolded, it has also become apparent that many jets show hardening of the spectrum in the optical/UV, either alone or apparently connecting to the X-ray flux (as in 3C~273).  In many cases, these UV and/or X-ray components cannot be consistent with IC/CMB, simply due to their spectral shape. The IC/CMB flux must be extremely Doppler boosted ($\delta\sim10-15$) to reach the observed flux levels in the UV through X-rays, which implies that the spectral index should be equal to that in the low-frequency radio.  This is often found to not be the case. Further, the UV and/or X-ray emission, under the IC/CMB model, comes from very low-energy electrons, which in turn implies extreme fluxes from the `peak' of the electron distribution at higher energies, which often over-predict either the X-rays (for UV components) or gamma-rays (for X-ray detected components).  Very hard second components have been found emerging in the UV with no or very low/soft X-ray emission, which is again not expected under an inverse Compton scenario -- for example the inner knots of PKS~2209$-$089 \citep{breiding2017}, and the outer knots of 3C~273 \citep{jester2006}.  A slight hardening of the X-ray spectrum in the inner knots of low-power jets like M87, 3C~346, and 3C~120 \citep{wilsonyang2002,worrall2004,harris2004} also suggests a second component, though much less dominant than found in powerful quasar jets.

As we present in this paper, the resolved jet in radio galaxy M84 has \emph{three} distinct components in the jet SED and is perhaps the best case of a low-power FR~I with a separate X-ray component.  As in \cite{breiding2017}, we propose the term `Multiple Spectral Component' or MSC jets for this broad class of objects where secondary (or tertiary) components are found, rather than the former 'anomalous X-ray jet' usage.  In general, these secondary and tertiary components are so far unexplained.

M84 is a very well-studied galaxy, being one of the four prominent ellipticals in the Virgo cluster ($z$=0.00339, $D$=18.5 Mpc), all of which are associated with hot X-ray gas \citep[$\sim$10$^7$\,K,][]{forman1984} within an intracluster gas of somewhat higher temperature ($\sim$3.5$\times10^{7}$\,K). M84 hosts a weak FR~I type radio jet, along with a counterjet of nearly the same length and luminosity, suggesting a source which is nearly in the plane of the sky. The black hole at the center of M84 is relatively massive for the general AGN population, but in keeping with the higher masses typically seen in FR~I hosting ellipticals, at $M_\mathrm{BH}=8.5^{+0.9}_{-0.8}\times 10^8M_\mathrm{\odot}$ \citep{walsh2010}. Optically the nucleus is classified as a LINER \citep{ho1997}, and the jet and surroundings have been extensively studied in the radio and X-rays.  \cite{laing_bridle} mapped the Faraday Rotation across the M84 jet, finding that the rotation is due to gas in front of the radio lobes.  The \emph{Chandra} images show that the radio lobes occupy lower-density `cavities' within the X-ray emitting gas \citep{fin2001,fin2008}, with the southern jet totally contained by the X-ray gas, as evidenced by the `shell' features seen in the radio \citep{laing2011}, while the northern jet is breaking free. X-ray emission coincident with the radio jet was first detected in \emph{Chandra} imaging by \cite{harris2002}; the source was subsequently included in two surveys of X-ray jets \citep{massaro2011,harwood2012}, but no previous detection of the jet outside of the radio and X-ray bands has been reported.

In this paper, we report the detection of M84 in the optical and UV in archival HST imaging, and in the sub-mm in archival ALMA observations.  We also analyze many archival radio observations with the Very Large Array (VLA), both for measuring the jet SED and looking for proper motions of the jet. In combination with a re-analysis of three separate \emph{Chandra} observations of M84, we investigate the broad-band SED of M84 for the first time, and examine possible theoretical explanations for the unusual three-component SED, including hadronic models.
 In this paper we use the spectral index convention $f_\nu\propto\nu^{-\alpha}$ and an angular distance scale of 89 pc/$''$ for M84.

\section{Methods}

\subsection{Hubble Space Telescope}

M84 has been observed by several of the instruments on HST, as part of
projects focused on science other than the radio jet.
We first noticed the jet in the far-UV (F225W) WFC3
imaging during an archival search for optical counterparts to resolved
jets detected by \emph{Chandra}.  The jet is most obvious in the UV
because the competition from galaxy light is considerably less (the
uncertainty in the galaxy subtraction is the reason for the larger
error bars with the optical filters as described below), however it is
detected in all the deep imaging we examined, summarized in
Table~\ref{table:hstdata}.

\begin{deluxetable*}{lcccclc}[t]
\tablecaption{\label{table:hstdata} HST Observations} \tablecolumns{7}
\tablewidth{0pt}
\tablehead{
Instrument & Filter & $\lambda_c$ & Freq. & Exposure & Date        & PID \\
           &        &   (\AA)       & (GHz) & Time     &             &  
}
\startdata
ACS/WFC    & F850LP &  9036 &  & 2x560s  & 2003 Jan 21  & 9401 \\
ACS/HRC    & F606W\tablenotemark{*} &  6060 &  & 3x782s  & 2002 Dec 17  & 9493 \\
ACS/WFC    & F475W  &  4745 &  & 2x375s  & 2003 Jan 21  & 9401 \\
WFC3/UVIS  & F336W  &  3354 &  & 4x600s  & 2009 Nov 24     & 11583    \\[-5pt]
           &        &       &  & 4x600s  & 2010 Mar 29     & 11583    \\
WFC3/UVS   & F225W  &  2359 &  & 8x600s  & 2009 Nov 23-24  & 11583    \\
\enddata
\tablenotetext{*}{Polarization Imaging}
\end{deluxetable*}

We first analyzed the ACS/F475W data to produce a reference image.
This filter was best suited to this purpose, since far more point
sources are detected in the optical than UV, while the blue filter
offers a slightly better PSF than the long-pass optical imaging. The
F475W data were taken in a single visit, and we used the AstroDrizzle
package to stack the two exposures into the final image using default
parameters and a 0.04$''$ final pixel scale. Given the low
redshift of M84, the galaxy is quite large in angular extent and
covers a large fraction of the HST detectors. To prepare a suitable
reference image against which to align the other HST data
(particularly the UV where the galaxy is barely detected), we modeled
and subtracted the galaxy using IRAF tasks \texttt{ellipse} and \texttt{bmodel}, and
rotated the image into a North-up orientation.

Both UV datasets consist of two orbits (in the case of the F336W
filter data, in two separate visits separated by 4 months).  In order
to best align the UV data to the F475W reference image, we ran
AstroDrizzle on the two epochs separately. We again used default
parameters other than a final pixel scale of 0.04$''$. These
individual images were then matched to the reference frame using the
large number of point sources in each.  This was accomplished using an
approach very similar to that presented in \citep{meyer2015} to align
multi-epoch HST imaging of 3C~273. Briefly, we first made a
rough alignment by hand-selecting 5-6 point sources in common in the
two images and calculating the best-fit 6-parameter linear
transformation. With a starting
rough match we were able to identify (again by eye) several more
sources in common between the UV and reference images, including
resolved background galaxies.  There were 18 in the F225W images and
57 in the F336W images. By selecting a region of pixels covering each
point source (about 1000 pixels per source), we found the optimal
shift to match the reference image source to the UV image (after
scaling the flux scale of the image to roughly match the reference
image) \citep[see ][]{meyer2015}. These shifts were used to iteratively
improve the 6-parameter linear transformation best fit until no more
improvement in the RMS of the shift values occurred. The final
RMS error on the background source positions was $\sim$0.2 pixels (8
mas) and $\sim$0.08 pixels (3 mas) using 14 and 14 final sources in the two
epochs of F225W imaging and 40 and 47 in the F336W imaging.  The final
numbers of background sources were after rejecting outliers, some of
which could be due to non-negligible motion of foreground stars over
the 6 years between the reference and UV imaging, or residual cosmic
ray artifacts in the drizzle images. For each UV filter, the two epochs
were combined using a simple averaging of the two images.

The ACS/F850LP data was matched to the reference image using nearly
the same procedure as described above. While galaxy subtraction was
not necessary prior to matching the UV data, the galaxy is extremely
dominant in the F850LP imaging and differences between the flux
profile in the F850LP/F475W non-subtracted images could affect the
calculation of optimal position shifts. We thus modeled the galaxy in
the initial F850LP drizzle image using \texttt{galfit} in order to
produce a subtracted image for both calculating the alignment and
producing the final rotated/matched image. The galaxy was well-fit
(final $\chi^2$ value of 0.231) with three sersic components with
effective radii of 840, 177, and 64 pixels and indices of 1.00,1.32,
and 0.041 respectively\footnote{The very low sersic indices are
  somewhat unusual, but we found that a single-sersic component fit in
  all wavelengths gives indices more typical of jet-hosting giant
  ellipticals (n=2$--$3), though with a worse overall fit to the
  data.}. The smaller two sersic components had both mode 2 and mode 4
fourier components, while the largest sersic index was slightly boxy
(boxyness parameter 0.0483). The F850LP imaging consists of a single
visit and was matched to the reference image with 58 initial reference
sources. The final RMS error in the registration is 0.084 pixels or 3
mas, using 26 final reference sources.

We also analyzed a relatively shallow
ACS/HRC F606W polarization observation to look for signs of the
polarization signature one might expect from synchrotron
emission. Unlike the deeper imaging, this data was simply stacked into
individual polarization filter images using AstroDrizzle and combined
using the standard formulae \citep{koz2004} to produce a total linear polarization
image. The World Coordinate System parameters of all final images were made consistent using alarge number of common background sources and IRAF tasks \texttt{ccmap} and \texttt{ccsetwcs}.

\begin{figure*}[ht]
\vspace{20pt}
\begin{center}
\includegraphics[width=4.5in]{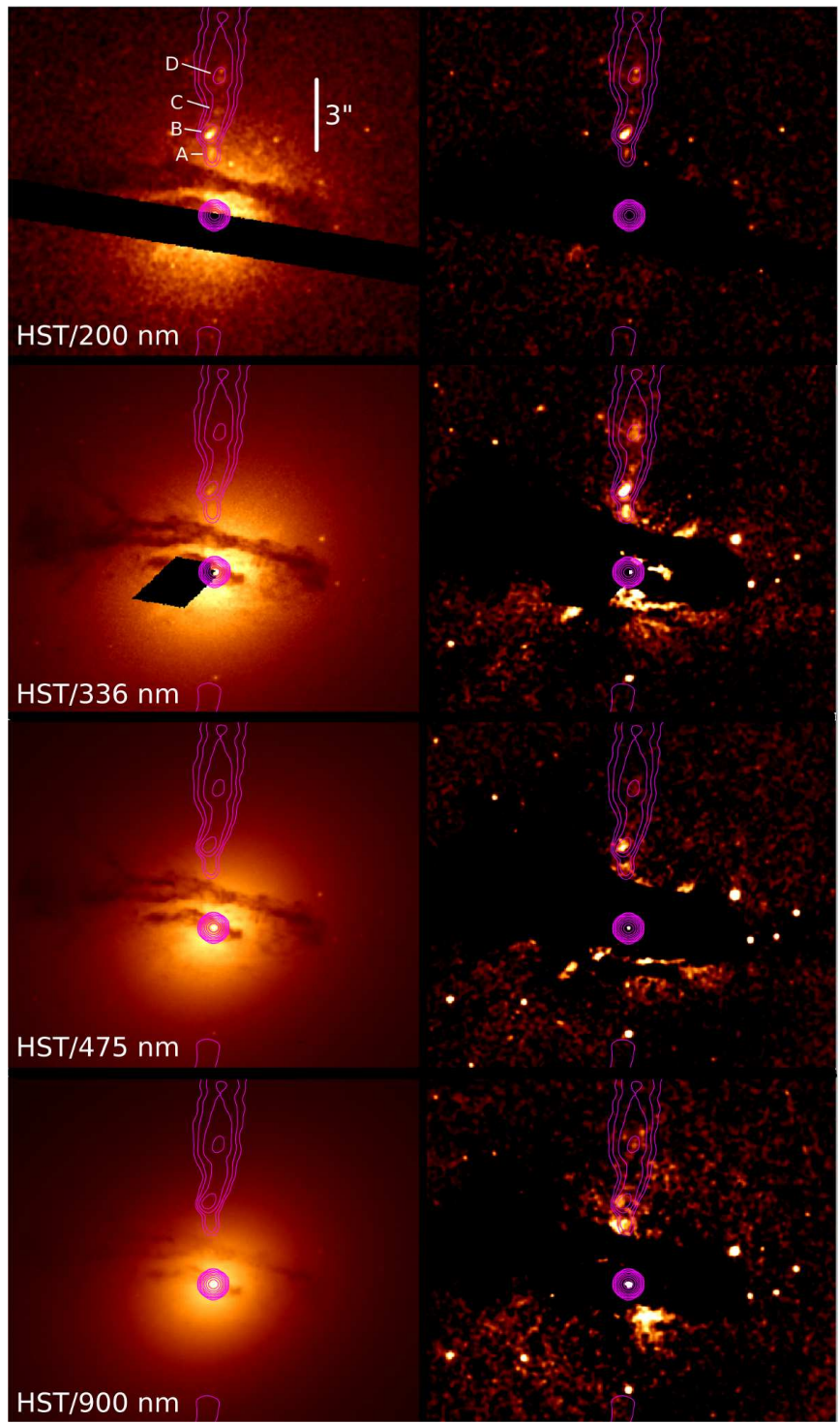}
\end{center}
\caption{\label{fig:hst} HST imaging of M84. In all cases the left
  image is prior to galaxy subtraction and the right is after
  (post-subtraction images have been smoothed with a 3-pixel gaussian
  kernel). From top to bottom, the camera/filter is WFC3/F225W,
  WFC3/F336W, ACS/WFC/F475W, and ACS/WFC/F850LP. Radio countours are
  from the C-band image, with the first level starting at 180 $\mu$Jy
  and subsequent levels by factors of 2. Four of the radio knots are
  clearly seen in the UV image as indicated at top, with the knots
  labeled A-D.  }
\end{figure*}

Prior to the final scaling of the images for analysis, the F225W,
F336W and F475W images were all galaxy-subtracted. The model in the
F475W filter was very similar to that of the F850LP image, with three
sersic components with effective radii of 850, 185, and 66 pixels and
indices of 1.00,1.29, and 0.047 respectively. The smaller two sersic
components had both mode 2 and mode 4 fourier components, while the
largest sersic index was slightly boxy (boxyness parameter
0.0382). The galaxy is much fainter in the UV images, so the fitting
is less complex. The F225W model required only one sersic component,
with radius 360 pixels and an index of 2.92, to reach a reduced
$\chi^2$ value of 0.918.  For the F336W model, three sersic components
were used, with radii of 616, 165, and 65 (indices of 1.09, 0.84,
0.59), but no other shape parameters, resulting in a reduced $\chi^2$
value of 0.104.

In Figure~\ref{fig:hst} we show the HST imaging, at left prior to
subtraction and at right after the subtraction, labeled with the
appropriate filters.  The overlaid contours are from the C-band VLA observation
described below. We note that the jet is clearly seen in the optical in the first three filters, while in the F850LP filter the morphology appears somewhat unclear.  While there is a statistically significant flux in the knot B region in this filter, it is not certain that the measured flux is correctly identified as coming from the synchrotron jet emission. However, this data point is not critical for any of the conclusions in this paper.

\begin{figure*}[t]
\vspace{20pt}
\begin{center}
\includegraphics[width=6.5in]{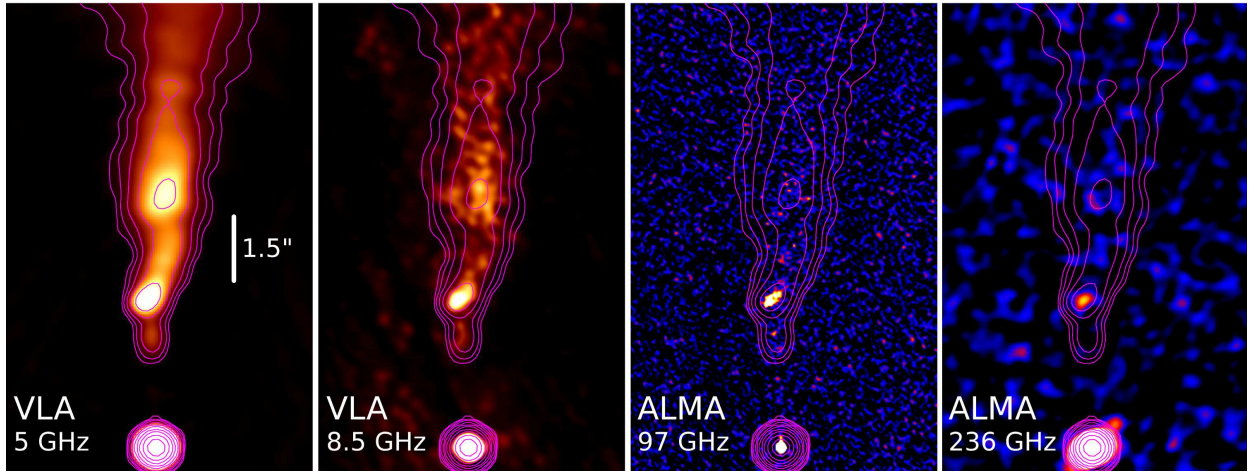}
\end{center}
\caption{\label{fig:vla_alma} From left to right, the VLA C-band and
  X-band images of M84, followed by the ALMA band 3 and band 6 images.
  All have C-band radio contours overlaid as in
  Figure~\ref{fig:hst}. For the ALMA images, the pixel scale is 5 and
  20 mas, respectively, and the images have been smoothed by a
  Gaussian of width 10 pixels.}
\end{figure*}

\subsection{Very Large Array}
\label{sec:vla_methods}
We analyzed archival high-resolution C- and X-band observations of M84
to best match the HST resolution for measuring individual knot
fluxes. The C-band observations were taken in A-configuration on 2000
November 18 as part of project AW0530, with 27 antennas, 200 MHz
bandwidth and 430 minutes on source over 10 hours producing excellent
UV coverage.  The data were reduced using CASA version 4.7.38348-REL
(r38335). After minor flagging of Radio Frequency Interference, a
standard calibration was applied, with 3C286 as the amplitude
calibrator and M84 itself serving as phase calibrator. Several rounds
of phase-only and a final round of amplitude and phase
self-calibration were applied before producing the final image. The
final image, shown at left in Figure~\ref{fig:vla_alma} has an RMS of
22 $\mu$Jy and a synthesized beam size of 0.43$''\times$0.41$''$.

The X-band data were taken in A-configuration on 1999 September 05 as
part of project AN8500, with 27 antennas, 200 MHz bandwidth and 10.4
minutes on source. The analysis was similar to the C-band data with 3C
286 again as amplitude calibrator.  The final self-calibrated image
has an RMS of 97 $\mu$Jy and synthesized beam size of 0.25$''\times$0.22$''$.

\subsection{Jet-to-counterjet Ratios}
\label{sec:cj_methods}
In addition to the above, we used lower-resolution archival VLA data
in L,C,X, and Ku band to study the jet-to-counterjet ratio. These were
an L-band, A-configuration data set from 2015 Jul 03 (project
15A-305), a C-band, C-configuration observation from 2000 Jun 04
(project AW0530), a X-band, C-configuration observation from 1998 Nov
28 (project AR0402) and a Ku-band, D-configuration observation from
2015 Dec 20 (project BH212). The L and Ku band observations were
calibrated using the JVLA pipeline, while the C and X data were
calibrated and flagged by hand.  All were imaged using clean in CASA
4.7.2 using standard procedures, including self-calibration.

The reason for using lower-resolution imaging for the counterjet
analysis was to avoid problems with missing flux due to the lack of
short baselines in the A-configuration, especially at higher
frequencies.  The resolution of the images was on the order of
1.3$''$, 3.5$''$, 2.1$''$, and 4.6$''$ and the largest angular scale
was on the order of 36$''$, 240$''$, 145$''$ and 97$''$ in L, C, X,
and Ku band respectively.

\subsection{VLA Proper Motions with WISE}

Two additional VLA observations were reduced to look for proper
motions of the knots. Both were C band, A-configuration to match the
resolution of the previous two high-resolution images. These were
taken in 1984 Dec 09 (project AH167) and 1988 Nov 23 (project
AW0228). These were reduced in the same manner as the epoch 2000 data,
and all were put on an aligned coordinate system with 20 mas pixel
scale. We attempted to use similar-resolution archival data at higher
frequencies, but none of the observations resulted in good detection
of the jet features, due to the short observing times.

To look for proper motions of the knots, we used the publicly
available Wavelet Image Segmentation and Evaluation (WISE) code
\citep{mertens2015_wise, mertens2016}.  The WISE code comprises three
main components. First, detection of structural information is
performed using the segmented wavelet decomposition method. This
algorithm provides a structural representation of astronomical images
with good sensitivity for identifying compact and marginally resolved
features and delivers a set of two-dimensional significant structural
patterns (SSP), which are identified at each scale of the wavelet
decomposition. Tracking of these SSP detected in multiple-epoch images
is performed with a multi-scale cross-correlation algorithm. It
combines structural information on different scales of the wavelet
decomposition and provides a robust and reliable cross-identification
of related SSP.

The final reported proper-motion values given in \S~\ref{sec:results_wise}
were obtained using a scale factor of 12 (for knot B) and 16 (for knot
A), as these were the scales which yielded a good detection of each
knot as a single component across multiple epochs.  Fainter knots C
and D in the jet were not consistently identified with a single
feature and could not be tracked with the limited epochs available.

\subsection{ALMA}

We searched through the available ALMA archival data on M84; none of
the observations available at the current time were taken with the
goal of imaging the jet. However, the brightest knot in the jet (knot B)
is detected in the band 3 and 6 imaging we present here.

The ALMA band 3 observation of M84 was taken as part of cycle 3
project 2015.1.01170.S where M84 served as a phase calibrator. The
nominal resolution and largest angular scale is 0.06$''$ and 0.55$''$,
respectively. We used the CASA pipeline version 4.5.1 to produce the
initial calibrated measurement set, while CASA version 4.7.1-REL
(r39339) was used for self-calibration and imaging with
\texttt{clean}.  The observation included 240 scans on M84 for a total
time on source of 72 minutes.  After several rounds of
self-calibration the final imaged was produced, with an RMS of 9 $\mu$Jy and
synthesized beam size of 0.06$''\times$0.05$''$.

For the ALMA band 6 image, we used an observation from cycle 2 project 2013.1.00828.S
(focused on measuring central black hole masses in AGN) where M84 was
one of the targets. The observations have nominal resolution and largest angular scale of 0.34$''$ and 1.74$''$,
respectively. We used the CASA pipeline version 4.3.1-REL (r32491) to
produce the initial calibrated measurement set, while CASA version
4.7.1-REL (r39339) was used for self-calibration and imaging with
\texttt{clean}.  The observation included 12 scans on M84 for a total
time on source of 49 minutes.  After several rounds of
self-calibration the final imaged was produced, with an RMS of 40
$\mu$Jy and synthesized beam size of 0.29$''\times$0.21$''$.

The VLA and ALMA images are shown in
Figure~\ref{fig:vla_alma}, with C-band contours overlaid.  While there is
some hint of increased brightness in the overall inner jet region in
the band 3 image, only knot B is clearly detected in both ALMA images.
In order to asess the effect of the minimum baseline on the largest
resolvable scale, we used the gaussian fit feature in CASA on knot B
in the higher-resolution X-band image, which gives a resolved size of
0.6$\pm$0.05$''\times$0.3$\pm$0.05$''$ for the knot.  The band 6 ALMA
image has a similar resolution to the X-band VLA image and finds a size
of 0.35$\pm$0.08$''\times$0.21$\pm$0.03$''$, suggesting the knot may
decrease in size with increasing frequency.  We thus do not expect a
significant loss in the flux measurement in the ALMA imaging due to
the short baseline limit of the observation, based on comparison to the largest angular scales, above.

\subsection{Chandra}

We analyzed three archival \emph{Chandra} observations of M84 taken with ACIS-S,
having IDs 803, 5908, and 6131 (taken on 2000 May 19, 2005 May 01, and
2005 Nov 07, respectively).  These observations were obtained to study
the interaction between the large-scale radio jet and the
inter-cluster medium
\citep{fin2001,fin2002,fin2008}. \cite{harris2002} published an
initial analysis of the M84 jet (finding two jet features) based on
the first epoch of \emph{Chandra} observations in 2000, with
subsequent analyses using all three epochs by \cite{massaro2011} and
\cite{harwood2012}, reporting three and two distinct jet features,
respectively. However, both of these later analyses of M84 were part
of a much larger study of several dozen jets, and a necessary level of
specificity in how the jet regions were selected is not available.  To
be certain that we measure the X-ray flux of the same regions we
identified in optical and radio imaging, we have re-analyzed the
M84 data.

\begin{figure*}[t]
\vspace{20pt}
\begin{center}
\includegraphics[width=6.5in]{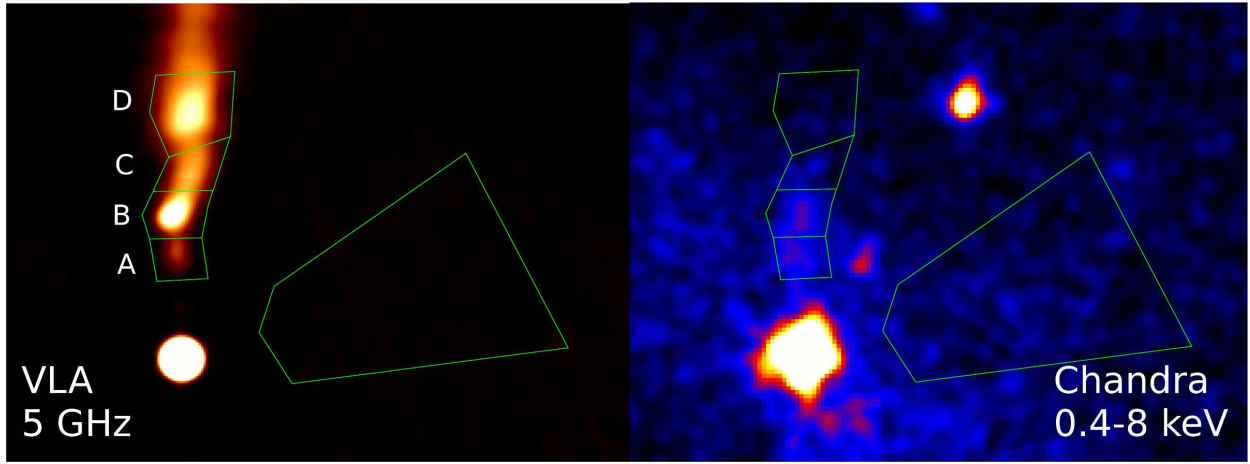}
\end{center}
\caption{\label{fig:chandra} At left, the 5 GHz VLA image of M84 with
  the \emph{Chandra} regions overlaid.  The \emph{Chandra} merged subpixel image
  (0.4-8 keV) with an exactly matching scale is shown at right, with
  the same regions overlaid. The large area to the right was used to
  estimate the background contribution, avoiding the non-jet related sources in the region.}
\end{figure*}

The data were analyzed using CIAO version 4.9. For each epoch, we
first ran the \texttt{chandra\_repro} script to generate level 2 event
files with the latest calibration applied. We made an initial energy
selection of 0.4$-$8 keV in keeping with standard practice of cutting
out the parts of the observing window with high background, and we
included only data from the S3 chip.  We then used the
\texttt{lc\_clean} script to remove background flares, resulting in
final exposure times of 23.545, 45.311, and 36.834 ks for the three
epochs, respectively.

We divided the jet into four regions, corresponding to knots A$-$D, as
shown in Figure~\ref{fig:chandra}. A background region was carefully chosen to approximate the level around the jet region (shown to the right of the
jet). The background is complicated by the fact that the environment
of the jet has a strong thermal signature. We extracted counts from
the same region in each epoch for analysis with \texttt{sherpa}.
Before analyzing the individual regions, we first fit the entire jet
area, to increase the statistics for measuring the spectral
parameters. We note here that no variability in the jet between epochs was observed, so we combined all observations in our analysis using standard methods. The model we used is a simple power law for the jet
components, with a thermal plasma emission model (APEC) for the background. The details of the fits are given in
Table~\ref{table:chandra}, where we give the (fixed) galactic
hydrogen value according to the mean value for the position in the
online \texttt{nH}
tool\footnote{https://heasarc.gsfc.nasa.gov/cgi-bin/Tools/w3nh/w3nh.pl},
the reduced statistic for the fit, the fixed redshift for the
background thermal model, and other important parameters for the APEC
and power law models.

\begin{deluxetable*}{cccccccccc}[t]
\tabletypesize{\scriptsize}
\tablecaption{\emph{Chandra} Spectral Fitting Results}
\centering
\tablehead{
   &                 & & \multicolumn{4}{c}{APEC Model}                        & \multicolumn{2}{c}{Power Law Model} & \\
  \cmidrule(lr){4-7}\cmidrule(lr){8-9}
  Component  & gal. $n_H$                  &    Red. $\chi^2$            &$z$  & $Z$\tablenotemark{a} & $kT$ & norm.           & $\Gamma$ & norm. &  $f_\mathrm{0.4-8,keV}$\tablenotemark{b} \\
           & $\times$10$^{22}$ cm$^{-2}$   &               & &                      & keV & $\times$10$^{-5}$ &          &  $\times$10$^{-7}$ & erg\,s$^{-1}$\,cm$^{-2}$
}
\startdata
Entire Jet & (0.026) & 0.65 & (0.003392) & (0.3) & 0.72$\pm$0.03 & 2.2$\pm$0.1 & 2.0$\pm$0.2 & 18.0$\pm$3.1 &  8.6$\times10^{-13}$\\
Knot A     & (0.026) & 0.75 & (0.003392) & (0.3) & (0.72)        & 2.2$\pm$0.1 & (2.0)       &  9.1$\pm$1.7 &  4.4$\times10^{-13}$\\
Knot B     & (0.026) & 0.71 & (0.003392) & (0.3) & (0.72)        & 2.2$\pm$0.1 & (2.0)       &  9.9$\pm$1.8 &  4.8$\times10^{-13}$\\
Knot C     & (0.026) & 0.73 & (0.003392) & (0.3) & (0.72)        & 2.3$\pm$0.1 & (2.0)       &  1.8$\pm$1.2 & $<$2.4$\times10^{-13}$  \\
Knot D     &         &      &            &       &               &             &             &              & $<$6.0$\times10^{-14}$ \\
\enddata
\tablecomments{Quantities in parentheses were fixed in the analysis. $\Gamma$ is the photon index. Normalizations are in standard CIAO units.}
\tablenotetext{a}{Metallicity for all species set to 0.3 solar values.}
\tablenotetext{b}{95\% Upper Limit Flux is reported for knots C and D based on all three epochs.}
\label{table:chandra}
\end{deluxetable*}

The results of the fit for knot C are given in the table, although the
normalization does not reach a reasonable significance threshold.
Likewise, knot D did not produce enough counts in the region to allow
a jet component to be included.  We thus calculated upper limits for
knots C and D using the ciao routine \texttt{aprates}, where
we set the PSF fraction to 1.0 and 0.0 in the individual knot and
background regions, respectively, and added the counts of all epochs
(and the total exposure time) to generate the 95\% upper limits given
in Table~\ref{table:chandra}.

Comparing to the two previous publications, our results are in
reasonable agreement.  The fluxes of the entire jet and four knots are
given in cgs units in Table~\ref{table:chandra}. The corresponding
best-fit monochromatic fluxes (at 1 keV) are 1.19$\pm$0.2, 0.60$\pm$0.11, and 0.66$\pm$0.12 nJy, for the
whole jet, and knots A and B respectively.  The knot D upper
limit is $<$0.08 nJy. In \cite{harwood2012}, the authors find a total
jet flux of 2.08 nJy with a somewhat harder photon index of
$\Gamma_x$=1.73$\pm$0.36, compared to our 1.19 nJy and
$\Gamma_x$=2.0$\pm$0.2, though it is not clear if any accounting for
the thermal background was made in the previous analysis.  In
\cite{massaro2011}, the three knots listed have distances from the
core of 2.7$''$, 3.5$''$, and 5.6$''$, which seems to agree with the positions of our
knots A, B, and D.  The fluxes given there were derived with an
assumed photon index of $\Gamma_x$=2, resulting in values of 0.23,
0.20, and 0.13 nJy, though the regions used were somewhat smaller than
used here (0.5$''$ diameter circular regions in all cases, about half
the size of our regions), which may explain the somewhat lower fluxes that they report.

\begin{deluxetable*}{llllllll}[t]
\tabletypesize{\scriptsize}
\tablecaption{\label{table:data}Properties and Multi-wavelength Fluxes for the knots of M84}
\centering
\tablehead{ Property & Observatory    & Frequency (GHz)     &  Flux Unit & knot A        & knot B        & knot C       & knot D}
\startdata
Distance (kpc)       &                &                     &            & 0.22          & 0.30          & 0.40         & 0.52   \\
Size (pc)            &                &                     &            & 43            & 50            & 39           & 140     \\
\hline 
$F_{\mathrm{L}}$     & VLA            & 1.51$\times10^{9}$  &  mJy       & \ldots        & 9.42          & \ldots       & 47         \\
$F_{\mathrm{C}}$     & VLA            & 4.86$\times10^{9}$  &  mJy       & 1.69          & 8.80          & 6.26         & 19.3       \\
$F_{\mathrm{X}}$     & VLA            & 8.46$\times10^{9}$  &  mJy       & 1.60          & 6.02          & 6.50         & 12.5       \\    
$F_{\mathrm{B3}}$    & ALMA           & 9.75$\times10^{10}$ &  mJy       & $<$0.95       & 1.03          & $<$0.95      & \ldots     \\
$F_{\mathrm{B6}}$    & ALMA           & 2.36$\times10^{11}$ &  mJy       & $<$0.4        & 0.30          & $<$0.66      & $<$0.53    \\
$F_{\mathrm{850LP}}$ & HST            & 3.33$\times10^{14}$ &  $\mu$Jy   & \ldots        & 15$\pm$6      & $<$21        & $<$42      \\
$F_{\mathrm{475W}}$  & HST            & 6.31$\times10^{14}$ &  $\mu$Jy   & \ldots        & 4.8$\pm$2.4   & $<$6         & $<$7.5     \\
$F_{\mathrm{336W}}$  & HST            & 8.92$\times10^{14}$ &  $\mu$Jy   & 1.2$\pm$0.2   & 2.7$\pm$0.4   & 1.1$\pm$0.3  & 2.9$\pm$0.4 \\
$F_{\mathrm{225W}}$  & HST            & 1.33$\times10^{15}$ &  $\mu$Jy   & 0.4$\pm$0.15  & 1.4$\pm$0.12  & 0.45$\pm$0.1 & 1.7$\pm$0.2 \\
$F_{\mathrm{1keV}}$  & \emph{Chandra} & 2.42$\times10^{17}$ &  nJy       & 0.60$\pm$0.11 & 0.66$\pm$0.12 & $<$0.32      & $<$0.08     \\
\enddata
\tablecomments{Distances from the core and sizes are converted using an angular distance scale of 89 pc/$''$. The sizes of the knots have been measured in the 1988 C-band radio image using a Gaussian fit, except for knot C which was measured by eye due to lack of compactness for the Gaussian fit. No errors are given for radio and ALMA fluxes, as the measured errors are smaller than the $\sim$5\% error assumed for the flux calibration. No flux measurement for knot A is reported for the two longer-wavelength HST filters due to contamination from galaxy residuals.}                    

\end{deluxetable*}
\subsection{Fermi/LAT Limits}

Fermi/LAT event and spacecraft data were extracted using a 10$^\circ$ region of interest (ROI), an energy range cut of 100 MeV-100 GeV, a zenith angle cut of 90$^\circ$, and the recommended event class and type for point source analysis.  The time cuts included all available Fermi data at the time of analysis with a corresponding mission elapse time (MET) range of 239557417-516452906.  Following the standard methodology for Fermi/LAT binned likelihood analysis, a binned counts map was created with 30 logarithmically spaced energy bins and 0.2  degree spatial bins.  An initial spatial and spectral model file was constructed with sources up to 10$^\circ$ outside the ROI using the publicly available \texttt{make3FGLxml.py} script.  This populates the model file with point and extended sources from the Fermi/LAT 3FGL catalog and an extended source catalog respectively. Additionally, the current galactic diffuse emission model, \texttt{gll\_iem\_v06.fits}, and recommended isotropic diffuse emission model for point source analysis, \texttt{iso\_P8R2\_SOURCE\_V6\_v06.txt}, was used. The livetime cube was computed using 0.025 steps in cos($\theta$) (where $\theta$ is the inclination with respect to the z-axis of the LAT) and 1$^\circ$ spatial binning. Then an all-sky exposure map was computed using the same energy binning as the counts map.  After obtaining a converged maximum likelihood fit for the model file, our target source of interest was added to the model file with a fixed photon index of 2.  After this, the upper limits in the five Fermi energy bands of 100 MeV-300 MeV, 300 MeV-1 GeV, 1 GeV-3 GeV, 3 GeV-10 GeV, and 10 GeV-100 GeV were computed by running the analysis tools separately on each data set with the appropriate energy range data cuts. These values are (in $\nu F_\nu$): 5.75$\times 10^{-14}$, 6.51$\times 10^{-14}$, 1.69$\times 10^{-13}$, 3.37$\times 10^{-13}$, 4.32$\times 10^{-13}$ erg~s$^{-1}$~cm$^{-2}$, respectively.

\begin{figure*}[!t]
\vspace{20pt}
\begin{center}
\includegraphics[width=6in]{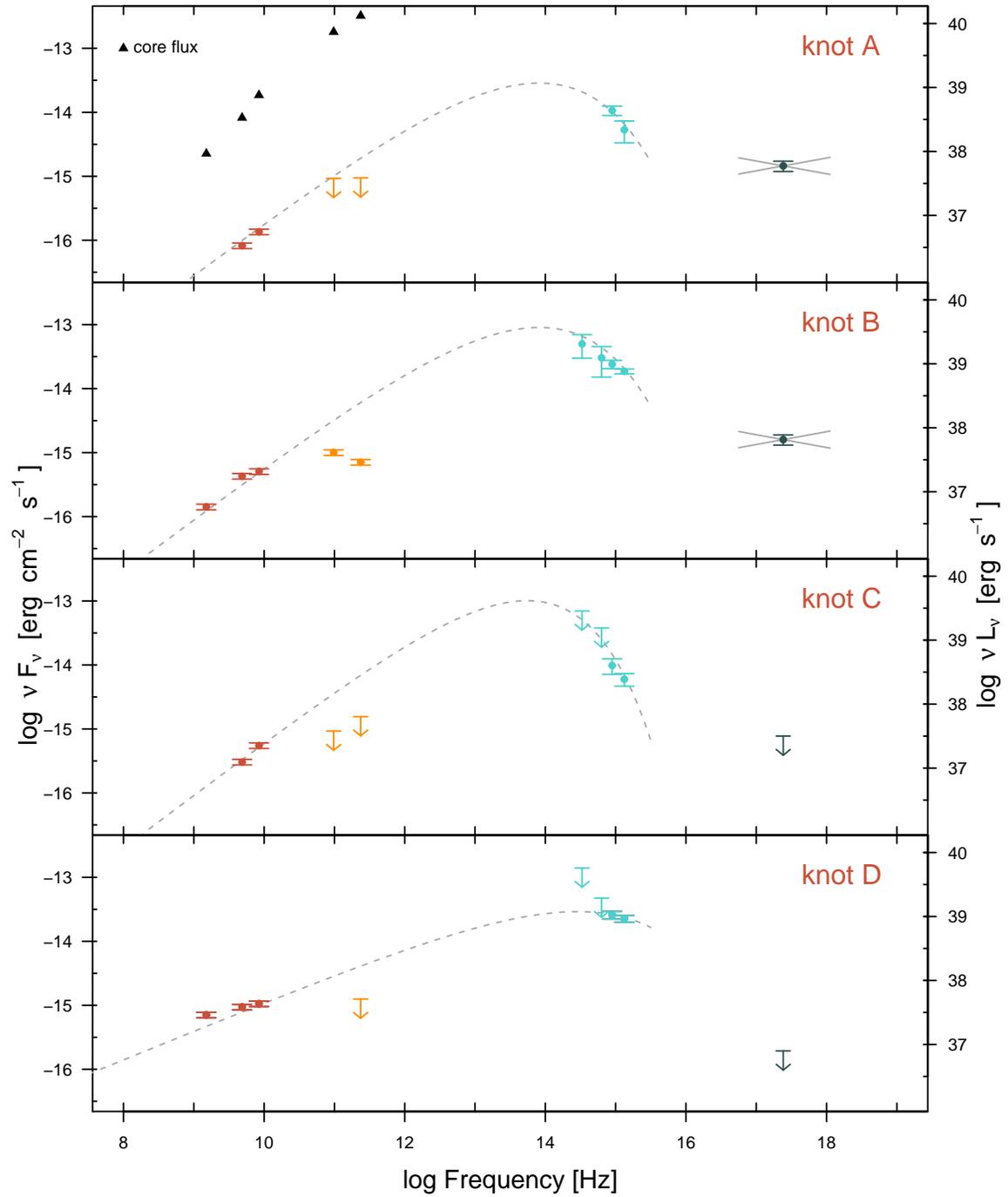}
\end{center}
\caption{\label{fig:sed_allknots} Radio through X-ray fluxes for knots
  A-D of M84.  In the top panel we also plot
  the radio fluxes of the core of the jet as black triangles for comparison (note that
  the ALMA fluxes show no turnover, unlike the ALMA points for knot
  B). Upper limits are shown as arrows. The gray dashed lines are phenomenological fits to the radio and optical data, to show how anomalous the ALMA data points are if we assume a single radio-optical synchrotron spectrum.}
\end{figure*}

\section{Results}

\subsection{The Unusual SED of M84 knots}

The resulting SEDs from radio to X-rays are shown in Figure~\ref{fig:sed_allknots}, where knots A through D are shown starting from the top panel (the corresponding data is given in Table~\ref{table:data}). The most striking feature of the SEDs is the turnover at about 100 GHz implied by the ALMA fluxes (for knot B) and upper limits (for knots A, C, and D).  The upper limits for the non-detected knots were calculated using the same region we used to measure the fluxes at other wavelengths. We show as a dashed line a phenomenological spectrum (power law plus exponential cutoff) connecting the VLA fluxes with the HST observations mainly to illustrate the anomaly of the ALMA data points, which are not compatible with the power-law spectrum. In the case of knot D, which is somewhat larger than the other knots, we do not report a band 3 upper limit because of the limit on the largest angular scale ($0.55''$) being smaller than the size of the knot ($\sim$1.5$''$), which means that the upper limit is likely to  be too low, considering the possibility of missing flux do to the array configuration. In the top panel of Figure~\ref{fig:sed_allknots} we also plot as black triangles the core flux in the radio and ALMA imaging. The core shows no sign of a similar turnover, so we can rule out a problem with the flux scaling as the origin of this spectral feature.    

The authors are not aware of any other published jet spectrum showing a similar turnover in the radio/sub-mm spectrum, though ALMA observations of jets are not yet extensive. However, previous ALMA observations of resolved jets usually show a continuous spectrum between radio and the optical, as in PKS~0637-752 \cite{meyer2017} and for several MSC jets in \cite{breiding2017}.  Given the observed turnover in M84, it is implied that the optical/UV comes from a second spectral component.  In addition, the flatness of the X-rays, and level, for knots A and B clearly indicate a third high-energy component which is not a continuation of the optical/UV.  We have ruled out reddening of the optical spectrum using a `dust map' image constructed by taking the ratio of the F336W and F850LP images.     Clearly, at least three spectral components are indicated by the SEDs.

\begin{figure}[t]
\vspace{20pt}
\begin{center}
\includegraphics[width=3.2in]{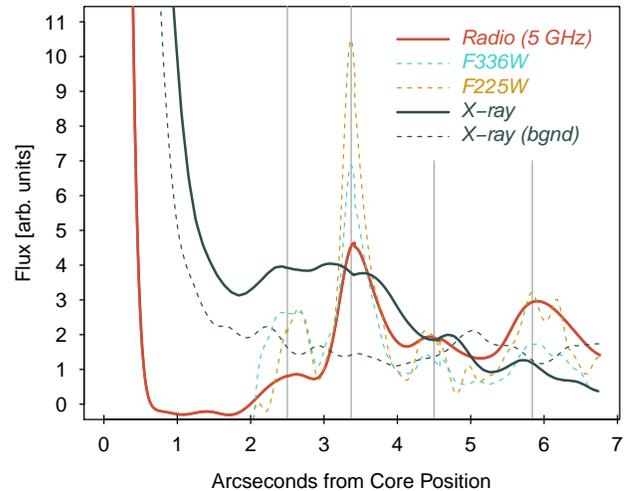}
\end{center}
\caption{\label{fig:contour} The 5 GHz radio, F225W and F336W HST UV, and \emph{Chandra} X-ray contour lines following the midline of the jet through knot centers, as defined by the 5 GHz radio image.  Flux units are arbitrary. We show an additional contour line taken from the X-ray image (dashed gray line) from the core through the center of the background region, to show that the X-ray level is essentially that of the background by $\sim$1.5$''$ from the core. In the jet X-ray contour, some contribution from the inner jet (interior to knot A) contributes to making the core seem to have a larger `wing'.}
\end{figure}

It is worth noting that the X-ray jet does not show the well-defined knots seen in the radio and UV.  This is apparent in the imaging shown in Figure~\ref{fig:chandra}, but is more clearly depicted in the contour line plot shown in Figure~\ref{fig:contour}.  Here we plot the radio flux from the 1988 C-band image as a red line, in comparison to the X-ray flux over the same contour line, following a path from the core out along the centers of each radio knot. Contours are also shown for the UV imaging as dashed lines. In all cases we take a `line average' perpendicular to the line running along the jet, extending out 0.14$''$, 0.2$''$, and 0.34$''$ in the radio, UV, and X-ray (approximately according to the PSF size), in order to yield a smoother contour line (the UV and X-ray images were also smoothed with a 3x3 gaussian kernal). The individual knots are not well-distinguished in the X-ray contour, though there is a strong drop in the flux level after knot B (around 4$''$) which seems roughly coincident with that in the radio and optical, taking into account the larger PSF of \emph{Chandra}.  The X-ray dominance (ratio of X-ray to radio flux) is clearly highest at the upstream end of the jet, which is similar to many of the powerful MSC quasar jets, such as 3C~273 \citep{marshall2001,jester2006} and PKS~0827+243 \citep{jorstad2006}.

\subsection{Optical and Radio Polarization}

\begin{figure*}[!hb]
\vspace{20pt}
\begin{center}
\includegraphics[width=6.5in]{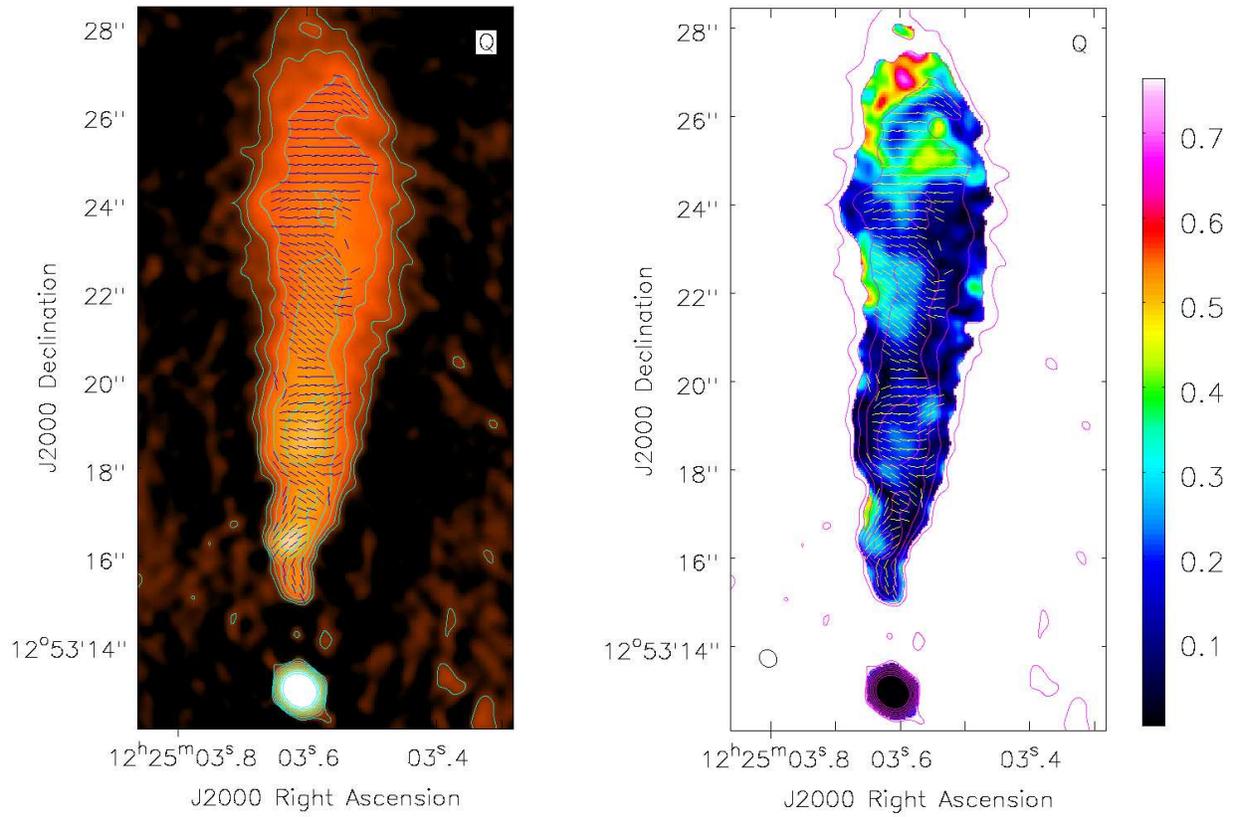}
\end{center}
\caption{\label{fig:vlapol} The total Intensity C-band (1988) image of the main M84 jet (left panel), and the Fractional linear polarization of the same (right panel, intensity scales shows fractional polarization level).  Overlaid on both are total-intensity contours in magenta starting at 0.1 mJy (spaced by factors of 2), and the magnetic field direction as black (white) lines. }
\end{figure*}

We examined the total linear polarization image from the ACS/HRC
imaging and found no hint of the jet features, with a 3$\sigma$ upper
limit on the polarization percentage of $<$8\% in the area of the
jet. However, the fractional polarization in the radio is not as low.
In the 1988 C-band image, shown in Figure~\ref{fig:vlapol},
linear polarization is detected throughout the jet.  Around the area of knot
B the highest polarization appears offset to the West but reaches
values on the order of 25\%, with similar levels in knot D, though somewhat lower in knots A and C ($\sim$10\%).

The low polarization in the optical is not necessarily unusual in the context of a synchrotron model (especially given the rather high upper limit), since synchrotron emission can still be produced in an environment with a `tangled' magnetic field which produces linear polarization that effectively cancels out.  However, it is somewhat unusual to find that a jet with reasonably high radio polarization lacks polarization in the optical \citep[see, e.g.,][]{perlman1999,perlman2006}.  Deeper optical polarization observations are necessary to get more stringent limits on the optical polarization degree, before any strong conclusions can be drawn about the nature of the optical emission.

\subsection{Counterjet Detection and Ratios}
We used the lower-resolution imaging in bands L, C, X, and Ku
(described in \S~\ref{sec:cj_methods}) in order to characterize the
jet-to-counterjet ratio.  Assuming that the jets are oriented 180
degrees from each other and have a consistent beaming pattern and
intrinsic power, this ratio contains information about the speed of
the jet and orientation to the line-of-sight.  In particular, the
ratio of the jet-to-counterjet flux can be expressed as

\begin{equation}
  R=(1+\beta cos\theta)^{m+\alpha}/(1-\beta cos\theta)^{m+\alpha},
\end{equation}

where $\beta$ is the speed of the jet in units of $c$, $\theta$ is the
angle of the jet to the line of sight, $\alpha$ is the radio spectral
index, and the index $m$ depends on the flow (in the case of moving
knots $m=3$ and for a continuous flow $m=2$). For M84, we take
$\alpha$=0.7 and $m$=3.

\begin{figure*}[t]
\vspace{20pt}
\begin{center}
\includegraphics[width=6.5in]{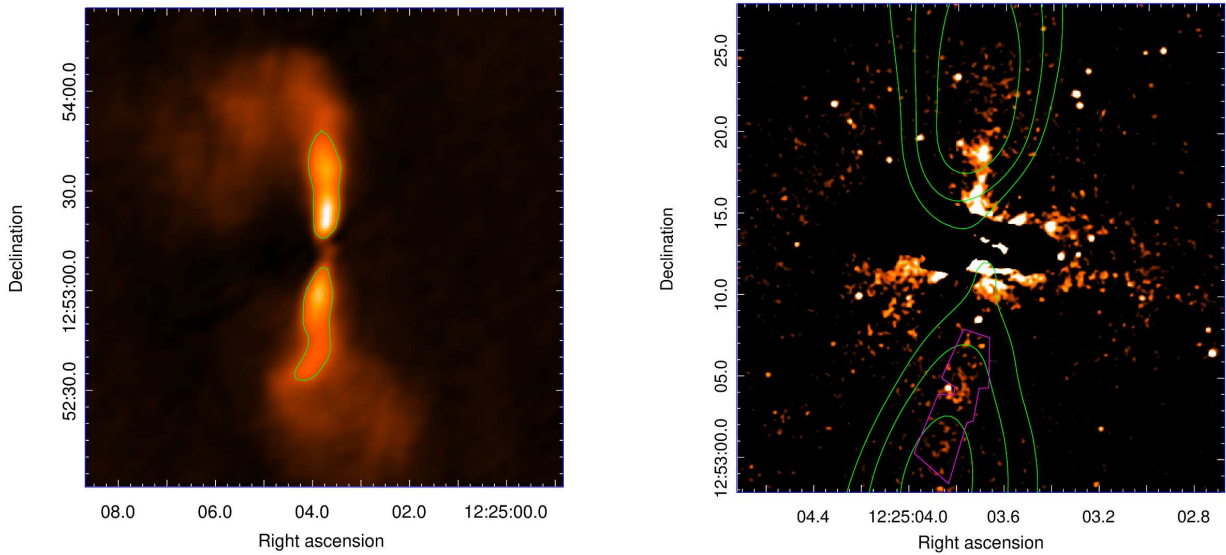}
\end{center}
\caption{\label{fig:counterjets} At left, the Ku-band D-configuration
  image of M84, with a contour line at 0.2 mJy overlaid (core has been subtracted).  The jet
  (upper) and counterjet (lower) regions were used to measure the
  jet-to-counterjet flux ratio $R$. At right, the WFC3/UVIS F336W
  image after galaxy subtraction and with appropriate scaling to show the faint
  counterjet.  Overlaid contours are from the Ku-band, D-configuration image,
  at levels of 0.1, 0.2, and 0.4 mJy. The magenta polygon region was
  used to measure the total flux of the feature.}
\end{figure*}

To measure $R$, we used a contour line at 0.2 mJy around the north and
south jet from the U-band image, which yields similarly sized smooth
regions surrounding the brightest part of each jet, as shown at left
in Figure~\ref{fig:counterjets}.  The purpose of making the
measurement at different frequencies was to look for evidence of a
change in the ratio with frequency, which might indicate a difference
in beaming.  However, we did not find such a trend, with values of 1.4, 1.6, 1.2, and 1.3 at L, X, X and Ku band, respectively.  We
thus take the mean value of 1.4$\pm$0.2 as representative of the radio
$R$ value.  Note that the errors on the flux measurement within the
contour are very small, less than 1\%, though some unaccounted error
comes from the exact choice of contour.  We verified with several
trials that areas of both larger and smaller extent yield
similar ratios to the above.

\subsection{Detection of UV Counterjet}
In addition to the easily-seen radio counterjet, we have also detected
the counterjet in our the F225W and F336W WFC3/UVIS imaging.  This is
shown at right in Figure~\ref{fig:counterjets} for the F336 image,
where contours from the Ku-band/D-configuration image are overlaid.
The faint over-density inside the magenta polygon region has a total
flux of 4.8 uJy in the F336W image, after taking into account the
slight over-subtraction of the background in the vicinity. The
standard deviation of the background level is 0.95 $uJy$, making this
an approximately 5$\sigma$ detection.  Similarly, in the F225W image,
using an identically sized and similarly placed region, we measure a flux of 2.1 $uJy$ compared to
the background standard deviation of 0.4$uJy$, again yielding an
approximately 5$\sigma$ detection. Using a region of comparable size
on the main jet, we measure fluxes of 12.3$\pm$1.6 and 4.8$\pm$0.5 uJy
for the F336W and F225W images, respectively.  The $R$ value in the UV
is thus consistently higher than in the radio at 2.6$\pm$0.6 and
2.3$\pm$0.5, respectively.  This could be explained if the optical/UV is more beamed than the radio, due to a spine/sheath structure in the jet. Such a structure is consistent with the slightly smaller size of the knots in the UV compared to the radio: a factor of 38\% for knot D and 30\% for knot B, where we used the gaussian fit tool in \texttt{casaviewer} and average the results for the two UV images (knots A and C did not give reliable gaussian fits in the UV images).

\subsection{Constraints from Proper Motions and Counterjet}
\label{sec:results_wise}
The results of the WISE proper motion analysis yielded a speed of 3.9$\pm$1.5 and 3.3$\pm$2.5 mas/yr for knots A and B,
respectively.  Results were consistent over a range of pixel scales
from 8-16, though the errors are relatively large.  Taking the knot A measurement
as the more constraining, this is equivalent to an apparent speed $\beta_\mathrm{app}$ of
1.1$\pm$0.4$c$.

\begin{figure}[t]
\vspace{20pt}
\begin{center}
\includegraphics[width=3.5in]{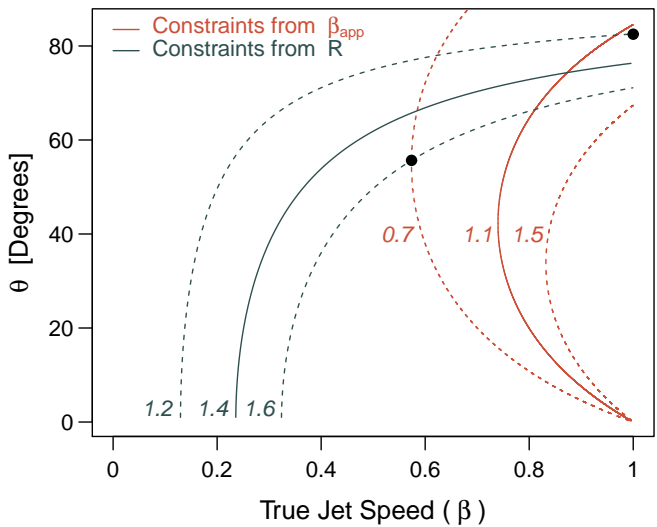}
\end{center}
\caption{\label{fig:constraints} Constraints in the $\theta-\beta$ plane based on the jet-to-counterjet ratio $R$=1.4$\pm$0.2 (dark gray lines) and proper motions $\beta_{\mathrm{app}}$=1.1$\pm$0.4 (orange lines).  The resulting best-fit angle to the line of sight is 74$^{+9}_{-18}$ degrees. The combination of limits leading to the lower and upper bounds on the angle are noted with black points.}
\end{figure}

Combining the proper motion measurement with the radio jet-to-counterjet
ratio yields a single value for the intrinsic speed of the jet as well
as the orientation angle.  This is shown in
Figure~\ref{fig:constraints}. Clearly, the data are consistent with
M84 having a relatively large angle to the line-of-sight.  Using the
bounds reported above on $\beta_\mathrm{app}$ and $R$, the resulting
angle is 74$^{+9}_{-18}$ degrees, where the lower bound is found from the maximum $R$ and minimum $\beta_\mathrm{app}$, and the upper bound from the lower limit of $R$ when the true jet speed is 1, as shown in Figure~\ref{fig:constraints}. It is also interesting that the low
value of $R$ precludes a value of $\beta_\mathrm{app}$ much larger
than we have observed. The angle of 74 degrees corresponds to a
$\beta$ value of 0.87, and a Lorentz factor $\Gamma$=2 (the lower
angle limit corresponds to $\Gamma$=1.2).

\begin{figure}[t]
\vspace{20pt}
\begin{center}
\includegraphics[width=3.5in]{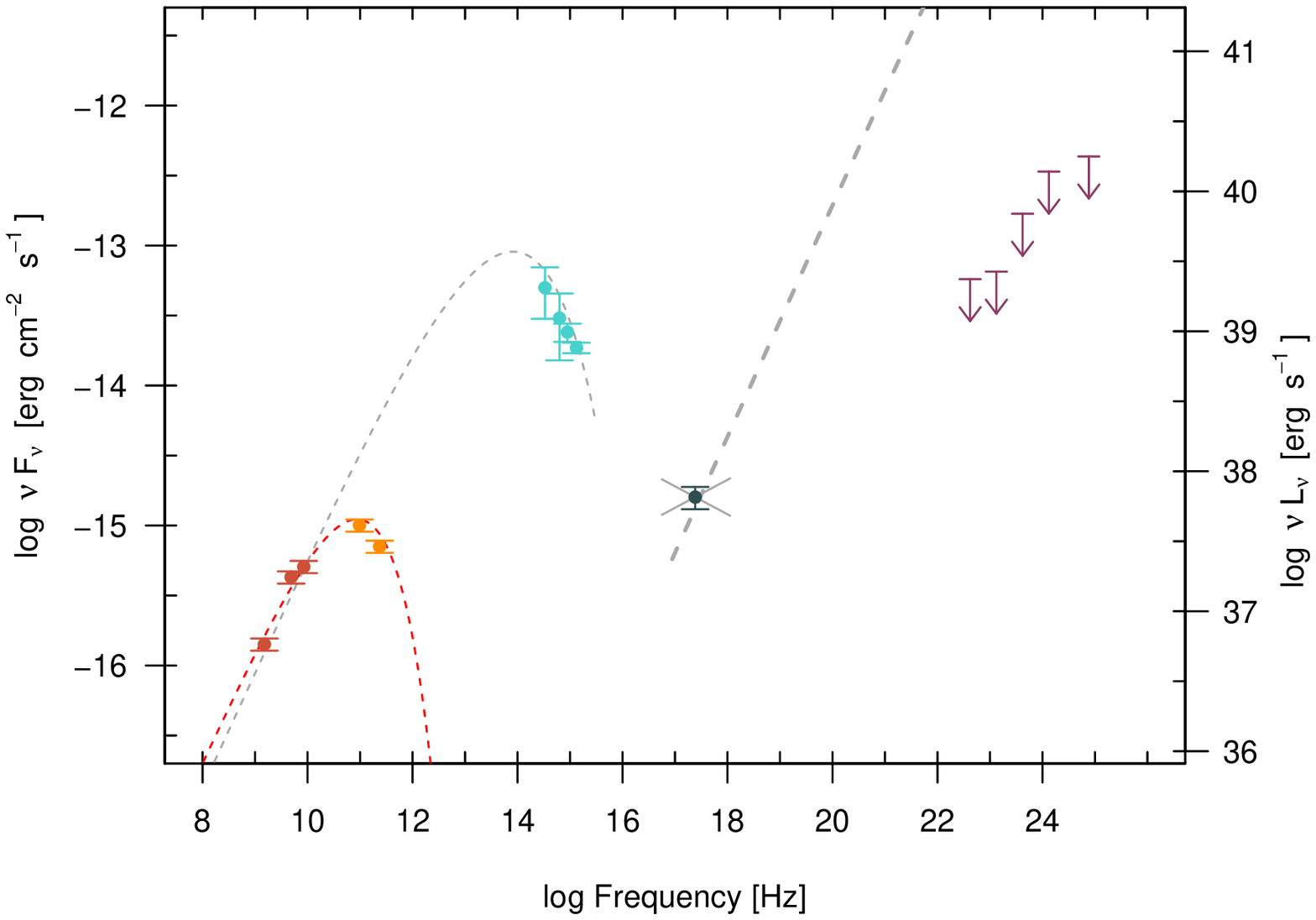}
\end{center}
\caption{\label{fig:sed_knotB} The SED of knot B, along with gamma-ray upper limits from \emph{Fermi}. The light dashed gray line fits the radio-optical (ignoring the ALMA data), while the red-dashed line fits the radio and ALMA data only. The X-rays are clearly too high and too hard to be connected to the optical component, however the spectral index is too soft for an IC/CMB model (thick dashed line at right). Further, an IC/CMB model would over-predict the Fermi limits, and require a magnetic field out of equipartition by three orders of magnitude. Only one IC/CMB curve is shown since they are virtually indistinguishable for either of the two synchrotron spectra shown over the range of frequency and flux shown here. Both severely violate all the Fermi upper limits.}
\end{figure}

\section{Interpretation of the SED}

\subsection{Star Formation}
One possibility for optical/UV emission in the vicinity of a relativistic jet is the presence of vigorous star formation induced by the jet impacting dense molecular clouds.  Such an origin would be consistent with the lack of optical polarization observed in M84. Direct observation of jet-induced star formation is rare, but it has been observed in a few low-redshift radio galaxies such as Centaurus A \citep{mould2000,salome2017}, 3C~285 \citep{salome2015}, and Minkowski's Object \citep{vanbreugel1985,lacy2017}. At higher redshifts, some radio galaxies show an alignment of optical emission with the radio jet axis which has also been taken as evidence for jet-induced star formation \citep[e.g.,][]{mccarthy1987}.  However, M84 would be a highly unusual case in comparison with previous cases, since we find that the UV knots perfectly correspond to the radio in location, and are actually somewhat more compact. This is similar to what has been found in other jets where there is no ambiguity as to the optical emission being synchrotron, due to spectral continuity and the large distance from the host galaxy \citep[e.g. PKS~0637-752;][]{meyer2017}. In contrast, previous cases of confirmed jet-induced star formation typically show filamentary structures (as in Centaurus A) or blobs which are displaced from the main radio emission, tending either to lie alongside or towards the jet terminus, and often having much larger angular extent, as in the high-redshift source 4C 41.17 \citep{bicknell2000}.  In these other cases, typical star formation rates are on the order of 1-100 $M_\odot$/yr. Using the basic scaling $\mathrm{SFR}_\mathrm{UV}=1.4\times 10^{-28}L_\mathrm{UV}$ \citep{rosa-gonzalez2002} with the far-UV luminosity of knot B ($L_B$= 6$\times$10$^{23}$\,erg~s$^{-1}$~Hz$^{-1}$) we find a star formation rate of $\sim10^{-4}$\,$M_\odot$/yr would be implied, before accounting for dust reddening, which might make this value 5-10 times higher. This is clearly still too low to be plausible when compared with the dynamical timescale of the jet -- knot B, for example, has taken approximately 3$\times 10^4$ years to arrive at its current location. Identical arguments would rule out star formation in the other knots as well, given their similar placement, size, and luminosity.
For these reasons we do not support star formation as a likely origin for the optical/UV emission in the M84 jet.

\subsection{Leptonic Models}
\subsubsection{Inverse Compton Models}
If we discount the low ALMA fluxes, a continuous synchrotron spectrum
can be fit to the radio-optical data, as shown with a dashed line in
Figures~\ref{fig:sed_allknots}~and~\ref{fig:sed_knotB}.  Clearly, given the observed ALMA fluxes and upper limits, these SEDs are not possible.  However, even assuming the single
radio-optical spectrum were correct, the X-rays are still clearly difficult to
account for, being too hard for the falling part of the synchrotron
spectrum.  As shown in Figure~\ref{fig:sed_knotB} for knot B, an IC/CMB model can be made to fit the X-ray flux level (thick dashed gray line),
but again the X-ray spectral index is off, only this time being far
too soft, and the model predicts a grossly unrealistic (and ruled out) level of gamma-ray emission, even with a turnover in the synchrotron spectrum at $\sim 10^{11}$~Hz.  Finally, the large angle and slow bulk motion implied by the
proper motions and counterjet measurements would require the IC/CMB
jet to be grossly out of equipartition.

\cite{harris2002} were the first to investigate in depth the possible emission mechanisms for the X-rays from the M84 jet, though without the benefit of the deeper follow-up observations taken in 2005. In that paper, the IC/CMB model is found to be unrealistic along the same lines we find here, and SSC is also ruled out as the photon density in the knot is so low that the predicted 1\,keV fluxes are four orders of magnitude below the observed values, assuming equipartition magnetic field strength. Allowing deviations from equipartition to match the observed X-ray fluxes would require a total jet power orders of magnitude larger than the Eddington luminosity, and would be unable to produce the flat spectrum in the X-rays -- thus SSC can be ruled out.
\subsubsection{Bremsstrahlung Model}
\cite{harris2002} also considered the possibility of thermal bremsstrahlung from relatively dense (5 cm$^{-3}$) hot gas, but favor a synchrotron interpretation due to the implied high pressure in the emitting region (far higher than the ambient medium, as well as the high temperature ($>$15\,keV) of the gas required under even adiabatic compression of the gas, which does not fit the observations.  We confirm that though the best-fit thermal bremsstrahlung model to the X-ray spectrum gives an acceptable fit (red. $\chi^2$ of 0.65) to the data, the temperature (2.2$\pm0.8$\,keV) is too low.	
 
 \begin{figure*}
 \centering
  \includegraphics[width=0.49\textwidth]{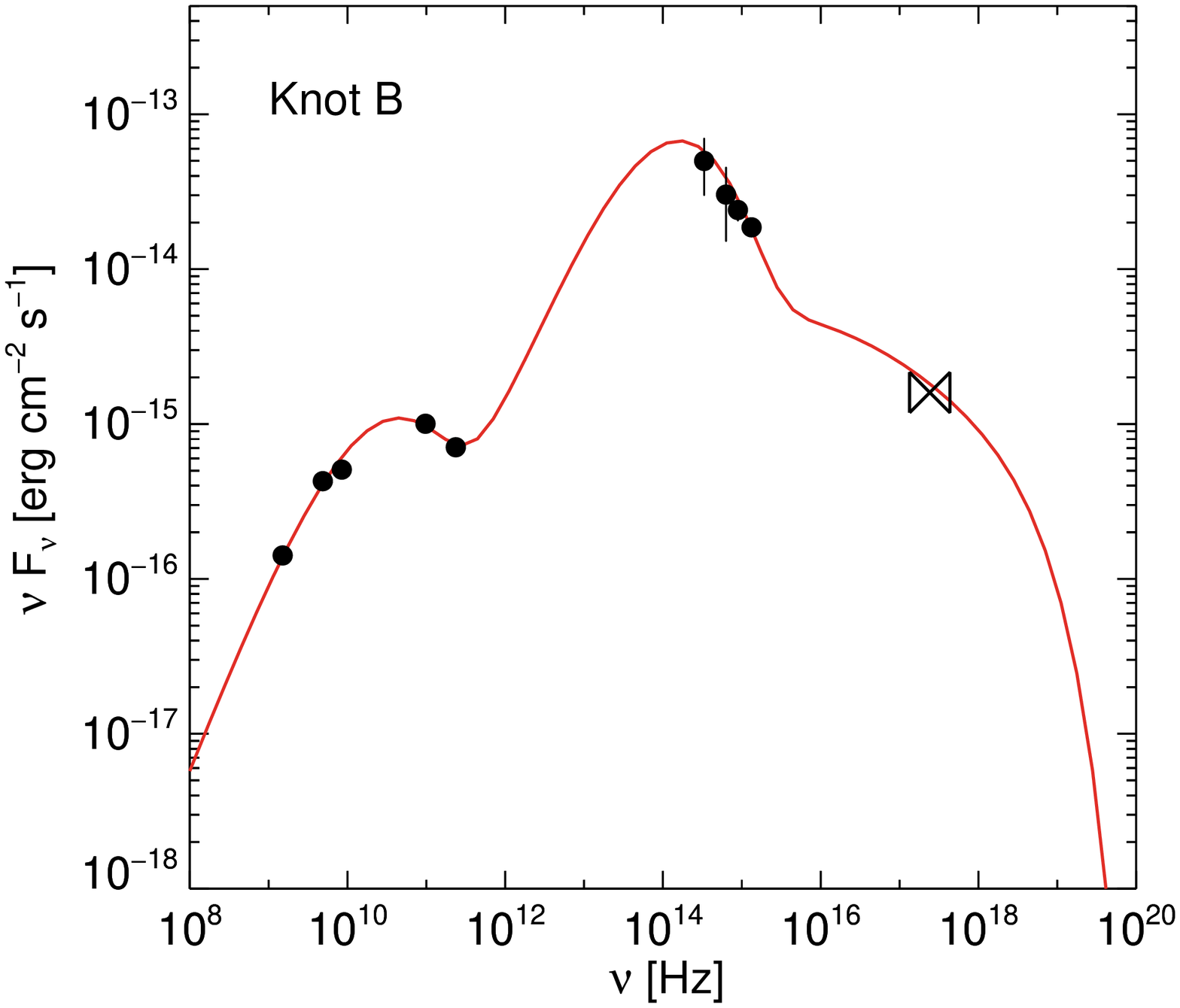}
  \includegraphics[width=0.49\textwidth]{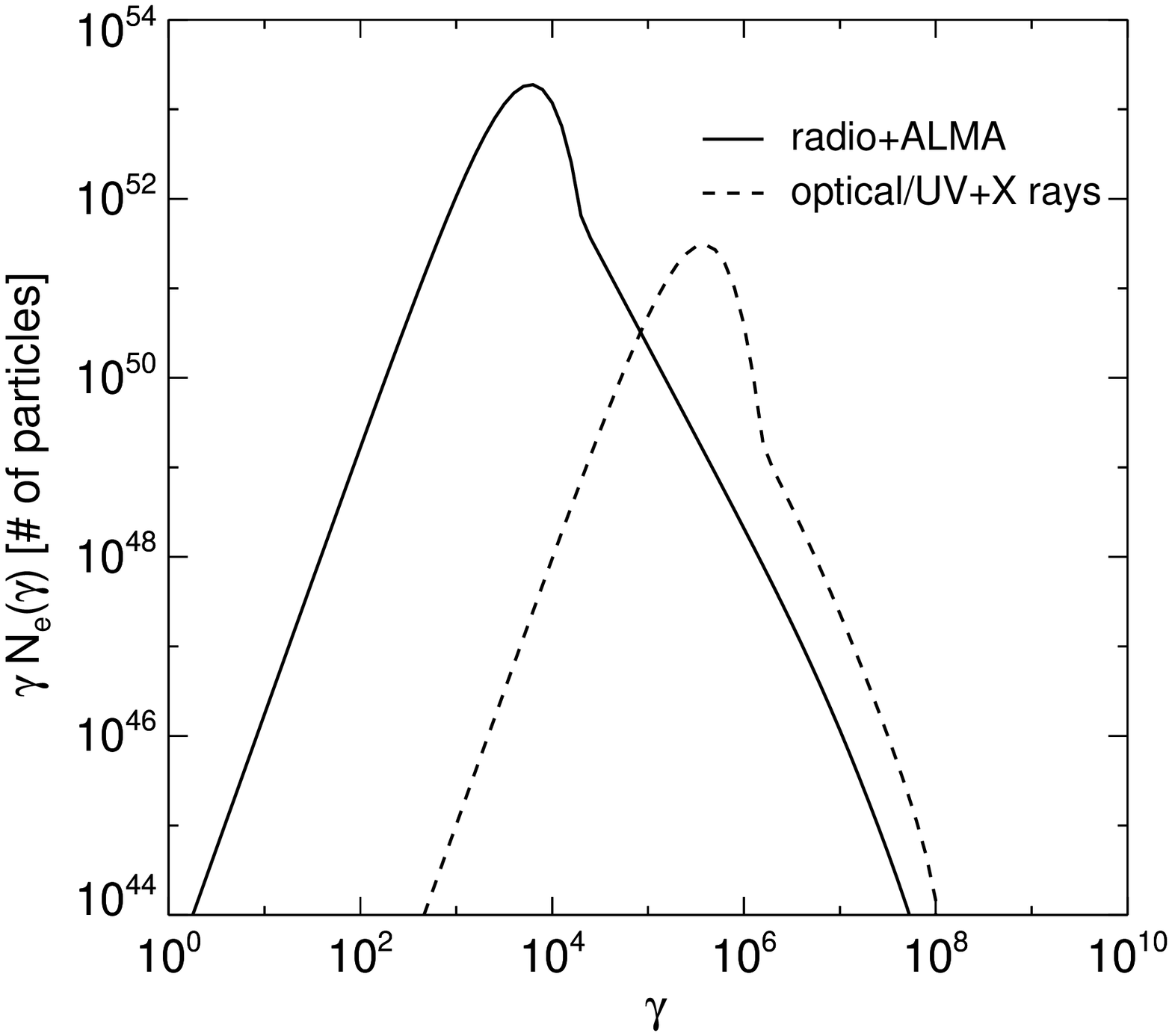}
  \caption{Left panel: two-component synchrotron fit (red solid line) to the SED of knot B (black symbols). Right panel: steady-state electron distributions invoked to explain the SED of knot B (right panel). We assumed that both can be described by a relativistic Maxwellian distribution with a power law. We find rough energy equipartition between the magnetic field and particles in knot B.}
  \label{fig:sed-elec}
 \end{figure*} 
\subsubsection{Two-component Synchrotron Model}
The problems of the inverse Compton models discussed in the previous section lead to the suggestion that the X-rays are also produced via synchrotron radiation from a second electron population residing in the knot \citep[e.g.][]{hardcastle2006,uchiyama2006,kataoka2008}. 

In this scenario, the radio and ALMA data are attributed to the synchrotron emission of the first electron population ($N_{e}^{(1)}$), while the second electron population ($N_{e}^{(2)}$) is responsible for the optical-to-X ray emission of the knot.  The different photon indices in the optical/UV and X-ray energy bands suggest that the underlying electron distribution is more complicated than a single power law. This could be a broken power law, although the hardening of the distribution above the break cannot be easily explained. Another possibility, which we adopt here, is that of a relativistic Maxwellian with ``temperature'' $\gamma_{th}\gg 1$ (in units of $m_e c^2$), smoothly connected to a power law from $\gamma_{nth}>\gamma_{th}$ \citep[e.g.][]{giannios_2009}. The particle distribution at injection can be written as:
\begin{eqnarray}
N_{e}(\gamma) = \left \{
\begin{tabular}{cc}
$C\frac{\gamma^2}{2\gamma_{th}^3}  e^{-\gamma/\gamma_{th}}$, & $\gamma \le \gamma_{nth}$ \\
$C \frac{\gamma_{nth}^2}{2\gamma_{th}^3} e^{-\gamma_{nth}/\gamma_{th}} \left(\frac{\gamma}{\gamma_{nth}}\right)^{-s}$, & $\gamma > \gamma_{nth}$ 
\end{tabular}
\right.
\label{eq:ne2}
\end{eqnarray}
We assume that both electron distributions are described by eq.~(\ref{eq:ne2}), although the radio and ALMA data can be explained by a single power-law electron distribution. All the model parameters of $N_e^{(2)}$ can be constrained by the data, except for the maximum Lorentz factor, whereas all the power-law parameters of $N_e^{(2)}$ remain unconstrained. These were assumed to be the same as those of the second electron population in order to reduce the number of free parameters. Moreover, we searched for parameter sets that result in (rough) energy equipartition between the magnetic field and the particles in the knot. 

Using the one-zone time-dependent numerical code of \cite{DMPR12}, we computed the steady-state photon and particle spectra within the two-component synchrotron scenario. Our fitting results are presented in Fig.~\ref{fig:sed-elec} and the parameter values are summarized in Table~\ref{tab:param-elec}. For these parameter values, we find that $u_{B}=4\times 10^{-10}$~erg cm$^{-3}$, $u_{e}^{(2)}=2 u_{e}^{(1)}=10^{-10}$~erg cm$^{-3}$, and that the power of an one-sided jet is $\sim 6\times 10^{42}$~erg s$^{-1}$.

Despite the success of the model in reproducing the multi-component SED for equipartition conditions in the knot, it relies on the \textit{ad hoc} assumption of two electron populations. There is no good explanation for the generation of relativistic Maxwellian distributions with temperatures that differ by almost four orders of magnitude, in a flow that is barely relativistic (see Sections 3.3 and 3.4).

\begin{table*}
\centering
\caption{Parameters of the two-component synchrotron model presented in Fig.~\ref{fig:sed-elec}.}
\begin{tabular}{cccc}
\hline
Symbol & Parameter & \multicolumn{2}{c}{Value} \\       
\hline
$B$ [mG] &  magnetic field strength &  \multicolumn{2}{c}{0.1} \\
$R$ [pc] &  knot size & \multicolumn{2}{c}{50} \\
$\delta$ & Doppler factor & \multicolumn{2}{c}{1} \\ 
\hline
                     &                       & $N_{e}^{(1)}$ & $N_{e}^{(2)}$ \\
$L_e$ [erg s$^{-1}$] &  injection luminosity & $8.8\times 10^{41}$ &$8.8\times 10^{41}$  \\
$\gamma_{th}$ [$m_e c^2$]& ``temperature'' of  Maxwellian & $2\times10^3$ & $1.3\times10^5$ \\ 
$\gamma_{nth}$& min. Lorentz factor of power law & $2\times 10^4$ & $1.3\times 10^6$\\
$\gamma_{\max}$ & max. Lorentz factor of power law & $10^8$ & $10^8$ \\
$s$ & power-law index & 3 & 3\\
\hline
\end{tabular}
\label{tab:param-elec}
\end{table*}
\subsection{Leptohadronic Model}
Models that consider the radiation produced by relativistic electrons and protons, the so-called leptohadronic models, pose an interesting alternative to the pure leptonic scenarios discussed in the previous section, f<or they can naturally produce multi-component SEDs (for blazars, see e.g. \citet{petropoulou15,petropoulou2017}; for large-scale jets, see e.g. \citet{aharonian2002, bhattacharyya16, kusunose2017}. Here, we focus on the SED of the brightest and, thus, best sampled knot of the jet (knot B). 

We model knot B\footnote{Here and in the next sections we focus on B simply because it is the best-sampled, but very similar modeling and similar results would apply to knot A as well.} as a spherical blob of radius $R=50$~pc containing a tangled magnetic field of strength $B$. Protons and electrons are assumed to be accelerated to relativistic energies and to be subsequently injected isotropically in the volume of the blob at a constant rate. Particles are also assumed to escape from the knot on a characteristic timescale of $R/c$. The accelerated particle distributions at injection are modeled as power laws between a minimum ($\gamma_{i, \min}$) and a maximum ($\gamma_{i, \max}$) Lorentz factor, namely  $N_{i}(\gamma)\propto \gamma^{-s_{i}}$ where $i=e,p$ stands for electrons and protons. 

At any time, the relativistic electron population is comprised of those that have undergone acceleration ({\sl primaries}) and those that have been produced by other processes ({\sl secondaries}). These include (i) the direct production through the Bethe-Heitler (BH) process $p \gamma \rightarrow p + e^{-} + e^{+}$, (ii) the decay of $\pi^{\pm}$ produced via the photopion process, i.e. $\pi^{+}\rightarrow \mu^{+}+\nu_{\mu}$, $\mu^{+}\rightarrow e^{+}+\bar{\nu}_{\mu}+\nu_{\rm e}$, and (iii) the photon-photon absorption $\gamma \gamma \rightarrow e^{+}+e^{-}$.  Secondary electrons also radiate via synchrotron and inverse Compton scattering processes. Due to the different energy distributions of primary and secondary electrons, the resulting SED will  be, in general, comprised of multiple components.  
\subsubsection{Analytical estimates}
The SED of knot B shows at least three components: the first (C1) peaks at $\nu_1\simeq 100$~GHz with a peak luminosity $L_{1,pk} \simeq 4\times 10^{37}$~erg s$^{-1}$, the second (C2) extends from sub-millimeter  to UV wavelengths with a likely peak at $\nu_2\simeq 10^{14}$~Hz and luminosity $L_{2,pk}\simeq 3\times 10^{39}$~erg s$^{-1}$, and the third (C3) emerges in the X-ray energy band. 

Components C1 and C2 can be naturally explained as synchrotron radiation of primary electrons and protons, respectively. In a magnetic field $B=1\ B_{mG}$~mG, the typical Lorentz factor of electrons and protons radiating at $\nu_1$ and $\nu_2$, respectively, is $\gamma_{e,pk} = \left(2\pi m_e c \nu_1/ B e\right)^{1/2}\simeq 6\times 10^3 B_{mG}^{-1/2} \nu_{1, 11}^{1/2}$ and $\gamma_{p,pk} = \left(2 \pi m_p c \nu_2/ B e\right)^{1/2}\simeq 8\times 10^6 B_{mG}^{-1/2} \nu_{2, 14}^{1/2}$. The spectral shapes of the two components also suggest that: $s_{e}\sim 2$, $\gamma_{e, \max}\simeq \gamma_{e,pk}$, $s_p\sim 4$, $\gamma_{p, \min}\simeq \gamma_{p, pk}$. Moreover, the maximum proton Lorentz factor cannot be arbitrarily large, as the synchrotron spectrum would extend to the X-rays, yielding a much softer spectral index than observed. Thus, $\gamma_{p,\max} \lesssim 10 \gamma_{p, pk}$. 

Protons with $\gamma_p > \gamma_{p, th}\sim m_e c^2/h \nu_2 \simeq 10^6 \nu^{-1}_{2,14}$ may directly produce secondary electrons with a wide range of Lorentz factors, i.e. from $\gamma_{e, \min}^{\rm (BH)} \sim \gamma_{p, th}$ to $\gamma_{e, \max}^{\rm (BH)}\sim 4 \gamma^2_{p, \max} h \nu_2 / m_e c^2 \simeq 10^{10} \nu^2_{2,14} B_{mG}^{-1}$ where we used $\gamma_{p, \max}=10 \gamma_{p, pk}$. Their distribution can be approximated by a power law with slope $p\simeq s_p$, while their synchrotron emission may emerge in the X-rays. The synchrotron spectrum can be approximately modeled as: 
\begin{eqnarray}
\nu L_{\rm BH}(\nu) \simeq \left(\beta-1\right) f_{\rm BH} L_p \left(\frac{\nu}{\nu_{\min}}\right)^{-\beta+1},
\label{eq:lum}
\end{eqnarray}
where $\beta=(p-1)/2$,  $\nu_{\min} \sim 10^{15}$~Hz, $\nu_{\max} \sim 10^{23}$~Hz, $L_p$ is the proton luminosity, and $f_{\rm BH}$ is a measure of the photo-pair production efficiency. This is defined as $f_{\rm BH} \equiv R \sigma_{\rm BH} n_2$ where $\sigma_{\rm BH}\approx 10^{-27}$~cm$^2$ is the average cross section, and $n_2\simeq 3 L_{2,pk}/4 \pi R^2 c h \nu_2$. Due to the very low number density of target photons photopair production is inefficient:
\begin{eqnarray}
 f_{\rm BH} \simeq  4 \times 10^{-6} \  L_{2,pk,39} \nu_{2, 14}^{-1} R_{pc}^{-1}.
 \label{eq:eff}
\end{eqnarray}
The observed X-ray luminosity, $L_X\simeq 8\times 10^{37}$~erg s$^{-1}$,  can be thus explained by synchrotron emission of Bethe-Heitler electrons (eq.~(\ref{eq:lum})), if the proton luminosity is sufficiently high, namely 
\begin{eqnarray}
L_{p} \sim 3\times 10^{45}\ R_{pc} \nu_{2, 14} L_{2,pk, 39}^{-1} \ \text{erg \ s}^{-1}.
\label{eq:Lp}
\end{eqnarray}
The synchrotron emission from secondary electrons produced in photopion interactions is expected to extend to even higher energies and be less luminous than the synchrotron radiation of Bethe-Heitler pairs \citep[see also][]{kusunose2017}. 
\subsubsection{Numerical modeling}
Here, we expand upon the scenario outlined in the previous section by performing detailed numerical calculations of the SED. For this purpose, we use the one-zone time-dependent leptohadronic numerical code of \cite{DMPR12} that includes all relevant radiative processes with the full expressions for the cross sections and injection rates of secondaries. 

\begin{figure}
 \centering
  \includegraphics[width=0.48\textwidth]{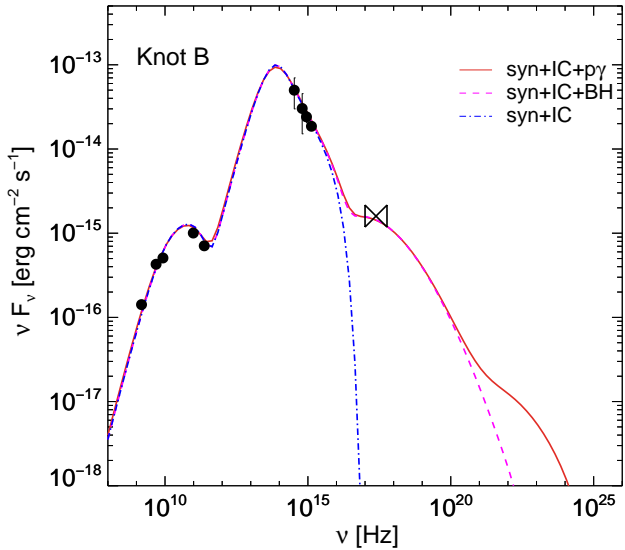}
  \caption{Leptohadronic fit (red solid line) to the SED of knot B (black symbols) where the radio peak is attributed to synchrotron from electrons, the optical peak to synchrotron from protons, and  the X-rays are attributed to the synchrotron radiation of secondary electrons from the Bethe-Heitler process. Overplotted (magenta dashed line) is the spectrum computed without photopion interactions (inverse Compton emission is included in the model although it is negligible here). The contribution of Bethe-Heitler pairs in X-rays is now evident. The spectrum obtained in the absence of photopion and photopair interactions is also shown for comparison (blue dash-dotted line).  The model parameters are summarized in Table~\ref{tab:param}. }
  \label{fig:sed}
 \end{figure}

We search for the best-fit parameter values by using as a starting point the rough estimates presented in the previous section. We consider as targets for photohadronic interactions only the photons produced in the knot B. We obtain the steady-state photon spectra after running the code for 610 light crossing times of the knot or $10^5$~yr. Our best-fit parameter values are summarized in Table~\ref{tab:param} and the resulting spectra are presented in Fig.~\ref{fig:sed}. The multi-component SED of knot B can be successfully described within the leptohadronic scenario. Because of the low number density of photons that are targets for photohadronic interactions, the proton luminosity must be about ten times larger than the Eddington luminosity of the source to explain the X-ray flux of knot B (Table~\ref{tab:param}). 

The energetic requirements could relax, if there were external optical photons with higher energy density than those produced internally in the knot. More specifically, the required proton luminosity could be reduced to $10^{46}$~erg s$^{-1}$, only if the energy density of optical photons was $\sim 10^{-9}$~erg cm$^{-3}$. Yet, the density of optical photons from both the core and the galaxy itself is many orders of magnitudes lower.

It is difficult at present to distinguish between the leptonic and hadronic scenarios with present capabilities. However, high-resolution imaging in the hard X-rays, with sufficient statistics to determine the hard X-ray spectrum of the knots, could help distinguish and inform jet models.  Given the upper limit on the optical polarization, future high-resolution X-ray polarization imaging of the knots would be useful to check the assumption of co-spatiality.

\begin{table}
\centering
\caption{Parameters of the leptohadronic model presented in Fig.~\ref{fig:sed}.}
\begin{tabular}{ccc}
\hline
Symbol & Parameter & Value \\       
\hline
$B$ [mG] &  magnetic field strength &  1 \\
$R$ [pc] &  knot size & 50 \\
$\delta$ & Doppler factor & 1 \\ 
$L_e$ [erg s$^{-1}$] &  electron luminosity (primary) & $9\times 10^{39}$ \\
$\gamma_{e, \min}$& min. electron Lorentz factor& $10^3$ \\ 
$\gamma_{e, \max}$& max. electron Lorentz factor& $5\times 10^3$\\ 
$s_{e}$ & electron power-law index & 2 \\
$L_p$ [erg s$^{-1}$] &  proton luminosity &  $2\times 10^{48}$$^{*}$\\
$\gamma_{p, \min}$& min. proton Lorentz factor& $3\times 10^6$\\ 
$\gamma_{p, \max}$& max. proton Lorentz factor& $5\times 10^7$\\ 
$s_{p}$& proton power-law index & 4.4\\
\hline
\end{tabular}\\
 Notes -- $^{*}$ The best-fit value of $L_p$ is ten times larger than the rough estimate (eq.~\ref{eq:Lp}) due to the simplifying assumptions entering the analytical calculation.
\label{tab:param}
\end{table}
 
\section{Conclusions}

In this paper we report the detection of a resolved optical/UV jet in
M84, extending $\sim$0.5\,kpc from the core, as well as a counterjet
detected at 5$\sigma$ in near and far-UV HST imaging. Combining
archival VLA, ALMA, HST, and \emph{Chandra} imaging of the jet in M84,
we have constructed a well-sampled SED for the kpc-scale jet in M84
for the first time.  We find evidence for a spectral turnover at
$\sim$100\,GHz which has not been previously observed in any
large-scale jet.  Along with the high optical flux and relatively hard
X-ray emission, this implies at least three distinct spectral
components in the M84 jet.  Archival HST polarization imaging in the
F606W band finds no optical polarization in the jet, with an upper
limit on the fractional linear polarization of 8\%.  In contrast, the
radio jet shows typical linear polarization fractions of 10-30\%.  We
rule out star formation as the origin of the optical/UV emission based
on the low UV luminosity. Leptonic inverse Compton models are ruled
out on the basis of requiring extreme departures from equipartition
and predicting a high gamma-ray flux which is not observed. We have
explored the possibility of a dual population of relativistic
electrons with relativistic Maxwellian distributions, coupled with a
power-law extension at high energies.  While such a model can fit the
observed SED of knot B in the jet, the model is ad-hoc and there is no
explanation of the extreme temperature difference between the
components.  We also explore a lepto-hadronic model in which the radio
and optical emission can be attributed to synchrotron emission from relativistic electrons and  protons, respectively, while the X-rays are due
to synchrotron emission from electron secondaries.  However, such a
model requires a total power in relativistic protons that is ten times
the Eddington limit.  At present we lack a comprehensive understanding of the physics of the emitting regions in the M84 jet.  These unusual features are not predicted by any known model of jet emission.

\acknowledgments
ETM and MG acknowledge NASA ADAP grant NNX15AE55G and NASA Fermi grant NNX15AU78G. MP acknowledges support by the L. Spitzer Postdoctoral Fellowship. MP would like to thank Dr. Dimitrakoudis and Prof. Mastichiadis for providing the numerical code. The National Radio Astronomy Observatory (NRAO) is a facility of the National Science Foundation operated under cooperative agreement by Associated Universities, Inc.
This paper makes use of the following ALMA data: ADS/JAO.ALMA\#2013.1.00828.S and ADS/JAO.ALMA\#2015.1.01170.S. ALMA is a partnership of ESO (representing its member states), NSF (USA) and NINS (Japan), together with NRC (Canada), MOST and ASIAA (Taiwan), and KASI (Republic of Korea), in cooperation with the Republic of Chile. The Joint ALMA Observatory is operated by ESO, AUI/NRAO and NAOJ.

\facilities{VLA, ALMA, HST, \emph{Chandra}, Fermi} 
\software{CASA, IDL, IRAF, CIAO, astropy}

\bibliographystyle{yahapj}
\bibliography{references}

\end{document}